\documentclass{article}
\usepackage{amsmath,amssymb,chicago,epsf,epsfig}

\newcommand{\rf}[1]{(\expandafter\ref{#1})}
\newcommand{\ct}[1]{\citeANP{#1}~[\citeyearNP{#1}]}

\newcommand{\lb}[1]{\expandafter\label{#1}}

\newcommand\mfk\mathfrak
\newcommand\mbb\mathbb
\newcommand\mtl\mathit
\newcommand\onm\operatorname
\newcommand{\bset}[2]{\bigl\{\,#1:#2\,\bigr\}}

\newcommand{\Hpl}{H^{\!\mbox{\scriptsize pl}}}
\newcommand{\omegapl}{\omega^{\mbox{\scriptsize pl}}}
\newcommand{\Jpl}{{J^{\mbox{\scriptsize pl}}}}
\newcommand{\Ppl}{P^{\mbox{\scriptsize pl}}}

\begin{document}\allowdisplaybreaks
\title{Reduction of the planar 4-vortex system\\ at zero momentum}
\author{George W. Patrick\\
Department of Mathematics and Statistics\\
University of Saskatchewan\\Saskatoon, Saskatchewan, 
S7N~5E6\\Canada
}
\date{January 2000}
\maketitle
\begin{abstract}
The system of four point vortices in the plane has relative equilibria
that behave  as composite  particles, in the  case  where three of the
vortices  have strength  $-\Gamma/3$   and  one of   the vortices  has
strength $\Gamma$.   These relative   equilibria occur at   nongeneric
momenta.    The   reduction  of this system,   at    those momenta, by
continuous and   then  discrete  symmetries, classifies  the  4-vortex
states which have been observed as  products of collisions of two such
composite particles.  In   this article I explicitly   calculate these
reductions, and show they are  qualitatively  identical one degree  of
freedom systems on a cylinder.  The flows on these reduced systems all
have one stable equilibrium and  one unstable equilibrium, and all the
orbits are  periodic  except  for  two homoclinic connections   to the
unstable equilibrium.
\end{abstract}

In the system of $N$ vortices in the plane the $n^{\mbox{\scriptsize
th}}$ vortex has location $z_n=x_n+iy_n\in \Ppl\equiv (\mbb C^2)^N$.
The Hamiltonian and symplectic form are
\begin{equation}\lb{5}
\Hpl\equiv-\frac1{4\pi}\sum_{m<n}\Gamma_n\Gamma_m\onm{ln} |z_n-z_m|^2,\quad
\omegapl\equiv\bigoplus_{n=1}^N\Gamma_n\omega_0,
\end{equation}
where 
$\omega_0(a,b)\equiv-\onm{Im}(a\bar{b})$ for complex numbers
$a,b\in\mbb C$. This system admits the symmetry group
$\mtl{SE}(2)=\{(e^{i\theta},a)\}$ of Euclidean symmetries acting
diagonally on each factor $\mbb C$ of $\Ppl$ by $(e^{i\theta},a)\cdot
z\equiv e^{i\theta}z+a$. A momentum mapping is
\begin{equation}\lb{7}
\Jpl\equiv-\sum_{n=1}^N\Gamma_n\left[\begin{array}{c}\frac12|z_n|^2\\iz_n,
\end{array}\right],\end{equation}
where $\mfk{se}(2)^*$ is identified with
$\mfk{se}(2)=\{(\dot\theta,\dot a)\}\cong\mbb R^3$ by the standard
inner product of $\mbb R^3$. For more details on the system of $N$
vortices on the sphere or plane,
see~\ct{KidambiRNewtonPK-1998.1},~\ct{PekarskySMarsdenJE-1998.1}, and
the references therein.

In~\ct{PatrickGW-1999.1} I have shown that the point of phase space
corresponding to a central vortex of strength $\Gamma$ at the origin
surrounded at distance $\alpha$ by three symmetrically placed outer
vortices of strength~$-\Gamma/3$ is a formally stable relative
equilibrium for this system; that is 
\begin{equation}\lb{4red-30}
z_1=\alpha,\qquad z_2=\alpha e^{2\pi i/3},\qquad z_3=\alpha e^{-2\pi i/3},
\qquad z_4=0
\end{equation}
is a formally stable relative equilibrium of the planar point vortex system
with $N=4$ and 
\begin{equation*}
\Gamma_1=\Gamma_2=\Gamma_3=-\frac\Gamma3,\quad \Gamma_4=\Gamma.
\end{equation*}
The momentum mapping is
equivariant for these vortex strengths because the sum of the vortex
strengths is zero. The relative equilibrium itself does not translate
but merely rotates with generator
$\dot\theta_e\equiv\Gamma/3\pi\alpha^2$, and by direct substitution
the momentum is $\mu_e\equiv(\Gamma\alpha^2/2,0)$. 
This momentum value is nongeneric, since it has isotropy group all of
$\mtl{SE}(2)$, and the position of this relative equilibrium is
unstable under small perturbation to nonzero momentum: as discussed
in~\ct{PatrickGW-1999.1}, the relative equilibria will move about on
the plane as a composite particle. To date these are the only
relative equilibria of point vortices which are both formally stable
and have nongeneric momentum. The particular system of four vortices
with the strengths $\Gamma_i$ is singled out by this fact.

When two such relative equilibria
collide the result may be localized 4-vortex states far from such
relative equilibria but each near to momentum~$\mu_e$, and these too
have been observed to move about on the plane as composite
particles. The phase space of classifying  these
states, since they are far from the relative equilibrium~\rf{4red-30},
yet still close to momentum~$\mu_e$, will be the reduced phase space
of the 4-vortex system at momentum~$\mu_e$. This phase space has
dimension $4\times 2-6=2$. Here I calculate this phase space and,
qualitatively, its associated Hamiltonian flow. One can find
other reductions on various point vortex systems, in the sphere and in
the plane, in~\ct{AdamsMRatiuTS-1988.1}, \ct{ArefHPomphreyN-1982},
\ct{EckhardtBArefH-1988}, and~\ct{PekarskySMarsdenJE-1998.1}. The
general issue of drifting relative equilibria at nongeneric momenta is
discussed at length in the sequence of articles~\ct{PatrickGW-1992.1},
\ct{PatrickGW-1993.1}, \ct{PatrickGW-1995.1}, and
\ct{PatrickGW-1998.1}.

To calculate the quotient
\begin{equation*}\pi_{\mu_e}:(\Jpl)^{-1}(\mu_e)\rightarrow
(\Ppl)_{\mu_e}\equiv(\Jpl)^{-1}(\mu_e)/\mtl{SE}(2),\end{equation*} one
may first translate the central vortex to the origin, whereupon, by~\rf{7}, the sum of the outer vortex positions must vanish. The
resulting 4-dimensional vector space $\mbb
C^2=\{(u_1+iv_1,u_2+iv_2)\}$ is spanned by the following orthonormal basis
$\{E_i\}$ of outer vortex positions:
\begin{equation*}
\epsfbox{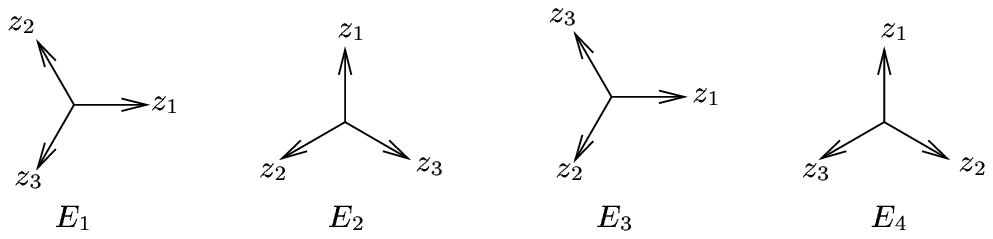}
\end{equation*}
or more specifically the column vectors of the matrix
\begin{equation*}
E\equiv\frac1{\sqrt 3}\left[\renewcommand{\arraystretch}{1.4}\begin{array}{cccc}
1&0&1&0\\
0&1&0&1\\
-\frac12&-\frac{\sqrt3}{2}&-\frac12&\frac{\sqrt3}{2}\\
\frac{\sqrt3}{2}&-\frac12&-\frac{\sqrt3}{2}&-\frac12\\
-\frac12&\frac{\sqrt3}{2}&-\frac12&-\frac{\sqrt3}{2}\\
-\frac{\sqrt3}{2}&-\frac12&\frac{\sqrt3}{2}&-\frac12
\end{array}\right].\end{equation*}
The $\mtl{SE}(2)$ symmetry then remaining on $\mbb C^2$ is
$\mtl{SO}(2)$ acting by diagonal clockwise rotation, and since the
above basis is orthonormal, the level set of $\Jpl$ becomes
\begin{equation*}
\bset{(u_1+iv_1,u_2+iv_2)}{
 {u_1}^2+{v_1}^2+{u_2}^2+{v_2}^2=3\alpha^2}.
\end{equation*}
The Hopf variables
\begin{alignat*}{2}
w_1&\equiv \frac2{3\alpha^2}(u_1v_2-u_2v_1),&\qquad w_3&\equiv\frac{-1}{3\alpha^2}\bigl(({u_1}^2+{v_1}^2)-({u_2}^2+{v_2}^2)\bigr),\\
 w_2&\equiv\frac2{3\alpha^2}(u_1u_2+v_1v_2),&\qquad
 w_4&\equiv\frac1{3\alpha^2}\bigl(({u_1}^2+{v_1}^2)+({u_2}^2+{v_2}^2)\bigr),
\end{alignat*}
with image the semi-algebraic set
\begin{equation*}
{w_1}^2+{w_2}^2+{w_3}^2-{w_4}^2=0,\quad w_4\ge 0.
\end{equation*}
are a quotient map for the diagonal $SO(2)$ action on this space, so
restricting these to the $w_4=1$ sphere gives the
quotient as
\begin{equation*}
({\Ppl})_{\mu_e}=\bset{(w_1,w_2,w_3)}{{w_1}^2+{w_2}^2+{w_3}^2=1}.
\end{equation*}
One partial section to the Hopf mapping, obtained by imposing $v_1=0$, is 
given by
\begin{equation*}
u_1=\frac{\sqrt3\alpha}2\sqrt{1-w_3},\quad v_1=0,\quad
u_2=\frac{\sqrt3\alpha}2\frac{w_2}{\sqrt{1-w_3}},\quad
v_2=\frac{\sqrt3\alpha}2\frac{w_1}{\sqrt{1-w_3}},
\end{equation*}
and a section to the map from $\Ppl$ to $\mbb C^2$ may be obtained by
setting $z_4=0$ and then
multiplying the vector $(u_1,v_1,u_2,v_2)$ by $E$ to obtain the triple
$(z_1,z_2,z_3)$. Using this section to pull
back the symplectic form on $\Ppl$ gives the reduced symplectic form
\begin{equation*}
\omega_{\mu_e}=\frac{\Gamma\alpha^2}4\omega_{S^2},\qquad
\omega_{S^2}(w)(\dot w_1,\dot w_2)\equiv-w\cdot (\dot w_1\times\dot w_2). 
\end{equation*}
To summarize, the symplectic reduced phase space is the 2-sphere with
symplectic form the constant multiple $\Gamma\alpha^2/4$ of the
standard symplectic form on the 2-sphere obtained by left-hand
reduction of the cotangent bundle of $\mtl{SO}(3)$.

The action of the discrete group $\mathfrak S_3$ corresponding to
permutation of the identical outer vortices descends to the reduced
space since it commutes with the action of $\mtl{SE}(2)$. Using the
above section this reduced discrete action is easily calculated to be
the restriction to the 2-sphere of the linear representation
$\sigma:\mathfrak S_3\rightarrow SO(3)$ given by

{\footnotesize\setlength\arraycolsep{2pt} \begin{equation*}\begin{cases}
\quad\sigma_{(12)}=\left[\begin{array}{ccc}
 \frac12&-\frac{\sqrt3}{2}&0\\-\frac{\sqrt3}2&-\frac12&0\\0&0&-1
 \end{array}\right],
\quad\sigma_{(13)}=\left[\begin{array}{ccc}
 \frac12&\frac{\sqrt3}{2}&0\\\frac{\sqrt3}2&-\frac12&0\\0&0&-1
 \end{array}\right],
\quad\sigma_{(23)}=\left[\begin{array}{ccc}
 -1&0&0\\0&1&0\\0&0&-1
 \end{array}\right],\\ \\
\quad\sigma_{(123)}=\left[\begin{array}{ccc}
 -\frac12&-\frac{\sqrt3}{2}&0\\\frac{\sqrt3}2&-\frac12&0\\0&0&1
 \end{array}\right],
\quad\sigma_{(132)}=\left[\begin{array}{ccc}
 -\frac12&\frac{\sqrt3}{2}&0\\-\frac{\sqrt3}2&-\frac12&0\\0&0&1
 \end{array}\right].
\end{cases}\end{equation*}}

The squares of the interparticle distances can be descended to the
reduced space, giving $\alpha^2/2$ times each of the 6 linear
functionals
\begin{gather*}
l_{12}\equiv-3(\sqrt3w_1-w_2-2),\quad l_{13}\equiv3(\sqrt3w_1+w_2+2),
\quad l_{23}\equiv-6(w_2-1)\\
 l_{41}\equiv2(w_2+1),\quad l_{42}\equiv-(\sqrt3w_1+w_2-2),\quad 
l_{43}\equiv\sqrt3w_1-w_2+2
\end{gather*}
and in terms of these the reduced Hamiltonian is
\begin{equation*}
H_{\mu_e}=\frac{\Gamma^2}{36\pi}\ln
 \left(\left(\frac{\alpha^2}2\right)^6
 \frac{(l_{41}l_{42}l_{43})^3}{l_{12}l_{13}l_{23}}\right).
\end{equation*}
The linear functionals $l_{ij}$ are independent of $w_3$ and are zero
for the two equilateral triangles circumscribing the unit circle in
the $(w_1,w_2)$ plane (see the left of Figure~\rf{4red-10}). The
Hamiltonian is finite except for the points in this plane where the
triangles touch the circle, where it is positive infinity on the
triangle that points downward, corresponding to collision states of
two outer vortices, and negative infinity on the triangle pointing
upward, corresponding to collisions states of an outer vortex with the
central vortex. The Hamiltonian has an amusing relationship with the
geometry of the collision triangles: it differs by a constant from the
function that is the logarithm of the product of the three distances
to the $-\infty$-collision triangle, cubed, divided by the product of
the three distances to the $+\infty$-collision triangle.

The equilibria, and hence the relative equilibria of the unreduced
system, may be found by calculating the critical points of
$H_{\mu_e}$. The south pole is the stable equilibrium corresponding
to the formally stable relative equilibrium~\rf{4red-30}, and the
north pole corresponds to the mirror image of this, obtained by
exchanging $z_2$ and $z_3$ in~\rf{4red-30}. There are six unstable
equilibria, namely the points of the orbit of the equilibrium $w_1=0$,
$w_2=\sqrt3-1$ under the action of $\mathfrak S_3$. One relative
equilibrium corresponding to these unstable equilibria is
\begin{equation}\lb{4red-41}
z_1=\alpha\sqrt[4]3,\quad
z_2=-\frac\alpha2\sqrt[4]3+\frac\alpha{2\sqrt2}(\sqrt 3-3)i,\quad
z_3=\bar z_2,\quad z_4=0.
\end{equation}
A graph of the reduced Hamiltonian as a function of $w_1$ and $w_2$
is shown on the right of Figure~\rf{4red-10}.
\begin{figure}\setlength{\unitlength}{1in}\centering\begin{picture}(4.5,1.4)
\put(0,.05){\epsfbox{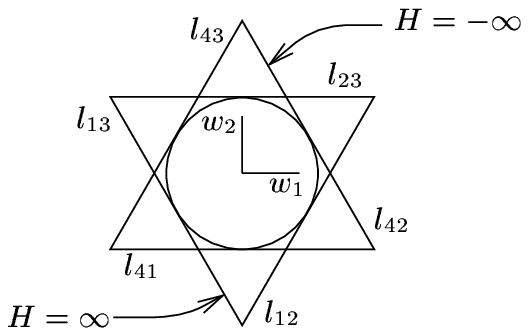}}
\put(1.5,-1.2){\epsfbox{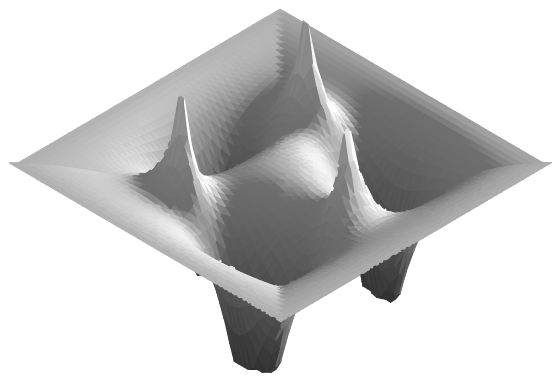}}
\end{picture}
\caption{\lb{4red-10}\protect\footnotesize\it Left: the level sets of
the linear functionals $l_{ij}$ are the triangles shown. Right: The
reduced Hamiltonian $H_{\mu_e}$; the sphere has been stereographically
projected from the south pole and then the plane compressed to a
square, so that the boundary of the square is the south pole. Clearly
visible are the $\pm\infty$ collision states and at the center the
stable equilibrium. Another stable equilibrium was, on the original
sphere, located at the south pole, so it occupies the boundary of the
square. Between the $+\infty$ collision ``mountains'' and the stable
equilibrium at the center are three saddle points and there are three
other saddle points between the $+\infty$ collision ``mountains'' and
the edge of the square.}
\end{figure}

The above is a complete analysis of the reduced space and the phase
portrait. However, to find the most succinct classification of the localized 4-vortex states, one should
also quotient by the action of $\mathfrak S_3$. Then the
unstable equilibria will collapse to a single point, the two stable
equilibria collapse to a single point, and there will be one $+\infty$
collision state and one $-\infty$ collision state. It is easy to
calculate that the stable equilibria have isotropy the cyclic group
$\mathfrak Z_3$, the collisions have isotropy the cyclic group
$\mathfrak Z_2$, and these are the only points of the reduced phase
space with isotropy.

To find the quotient of the reduced space by $\mathfrak S_3$ one can
find its image under sufficiently many functions invariant under the
action of $\mathfrak S_3$. By averaging up to degree~4 polynomials in
$w_1,w_2,w_3$, one finds the invariant polynomials
\begin{alignat*}{2}
&p_1\equiv{w_3}^2,&\qquad&p_2\equiv w_2(3 w_1^2-w_2^2),\\
&p_3\equiv w_1w_3(3{w_2}^2-{w_1}^2),&\qquad&p_4\equiv{w_1}^2+{w_2}^2+{w_3}^2
\end{alignat*}
and the relations
\begin{equation}\lb{4red-31}
p_1\bigl((p_1-p_4)^3+{p_2}^2\bigr)+{p_3}^2=0,\quad 0\le p_1\le p_4.
\end{equation}
The map $(w_1,w_2,w_3)\mapsto (p_1,p_2,p_3,p_4)$ can be seen as the
composition of simpler steps, namely
\begin{equation*}
 \tilde p_1\equiv w_3,\quad \tilde p_2+i\tilde p_3\equiv(-w_2+iw_1)^3,\quad
\tilde p_4\equiv{w_1}^2+{w_2}^2+{w_3}^2,
\end{equation*}
followed by
\begin{equation*}
p_1=\tilde p_1{}^2,\quad p_2=\tilde p_2,\quad p_3=\tilde p_1\tilde p_3,\quad
p_4=\tilde p_4,
\end{equation*}
and hence it is easily seen that $(w_1,w_2,w_3)\mapsto
(p_1,p_2,p_3,p_4)$, with image the semi-algebraic set~\rf{4red-31}, is
a quotient map for the action of $\mathfrak S_3$ on $\mbb
C^2$. Thus~\rf{4red-31} restricts to a quotient map of the reduced
space to the semi-algebraic set
\begin{equation}\lb{4red-32}
p_1\bigl((p_1-1)^3+{p_2}^2\bigr)+{p_3}^2=0,\quad 0\le p_1\le1.
\end{equation}
\begin{figure}[t]\setlength{\unitlength}{1in}\centering\begin{picture}(4.5,1.4)
\put(-.65,-.75){\epsfbox{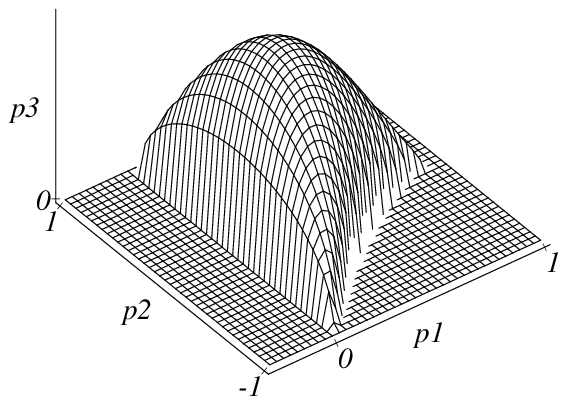}}
\put(1.85,-.75){\epsfbox{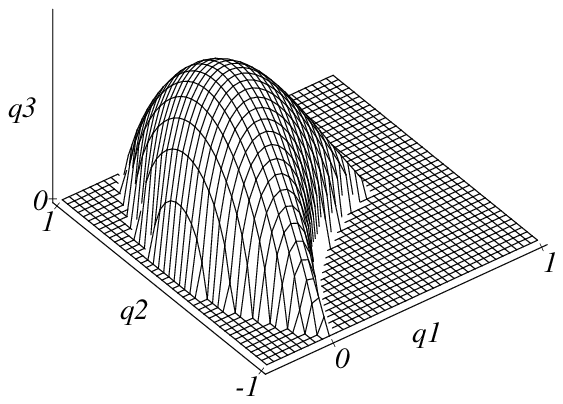}}
\end{picture}
\caption{\lb{4red-35}\protect\footnotesize\it Left: top half of the
semi-algebraic set~\rf{4red-32}. The $+\infty$ collision state is to the
front and the $-\infty$ collision state is behind and obscured; the
singularity at $(p_1,p_2,p_3)=(1,0,0)$ is the relative
equilibrium~\rf{4red-30}. Right: same as left but deformed according
to~\rf{4red-37}. The resulting surface may be radially projected to
the sphere and then stereographically projected to the plane from the
point $(q_1,q_2,q_3)=(0,-1,0)$.}
\end{figure}
The Hamiltonian this reduced space is
\begin{equation*}
H_{\mu_e}=\frac{\Gamma^2}{36\pi}\ln\left(\frac{\alpha^{12}}{2^43^3}
 \frac{(1+3p_1-p_2)^3}{1+3p_1+p_2}\right).
\end{equation*}
\begin{figure}[p]\setlength{\unitlength}{1in}\centering
\begin{picture}(4.5,3.)
\put(.1,-.1){\epsfbox{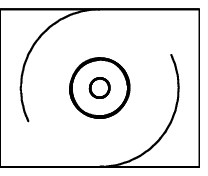}}
\put(.1,.7){\epsfbox{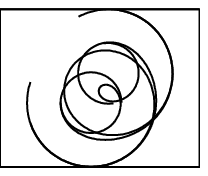}}
\put(.1,1.5){\epsfbox{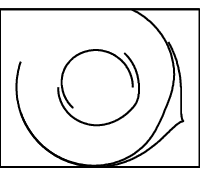}}
\put(.1,2.3){\epsfbox{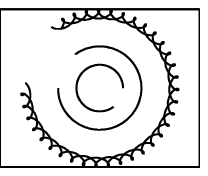}}
\put(1.6,.1){\epsfbox{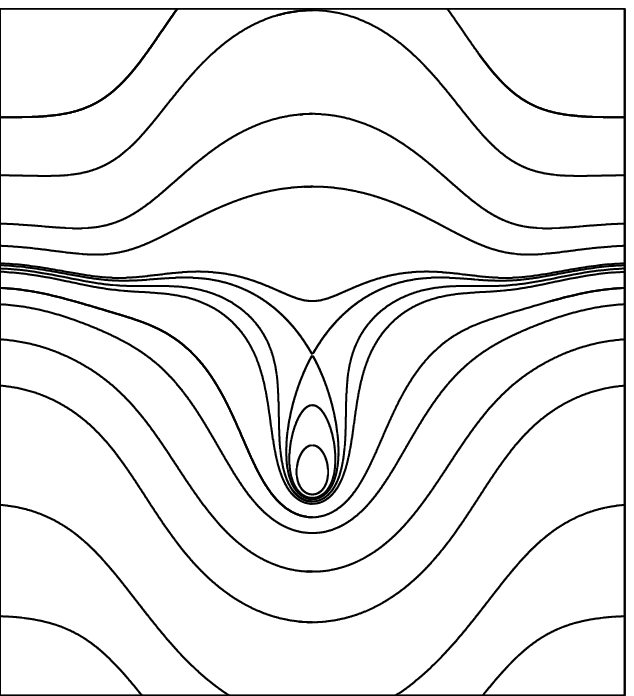}}
\end{picture}
\caption{\lb{4red-40}\protect\footnotesize\it Right: the result of
projecting, by~\rf{4red-50}, simulations of the 4-vortex system to the
toral reduced space as described. The horizontal variable is
$\theta_{\mu_e}$ and the two vertical sides of the phase portrait
should be identified. Left top to bottom: motions of the vortices on
the plane near the $+\infty$ collision, the top half of the homoclinic
orbit, the bottom half of the homoclinic orbit, and the $-\infty$
collision state. The pictures on the left are
reconstructions of orbits on the right. Left, 
the third graph from the top: the motion of the vortices is a close
pair for strengths $-1$ orbiting two vortices, one of strength $3$ and
the other of strength $-1$, and periodically the vortex on the inside
is exchanged with one of the vortices on the outside; this corresponds to the
lower half of the homoclinic orbit on the right. Left, second graph from the top:
the same configuration but the two outer vortices making up the close
pair exchange positions; this corresponds to the upper half of the homoclinic
orbit on the right.}
\end{figure}
\begin{figure}[p]\setlength{\unitlength}{1in}\centering
\begin{picture}(4,1.8)\put(.1,-.1){\epsfbox{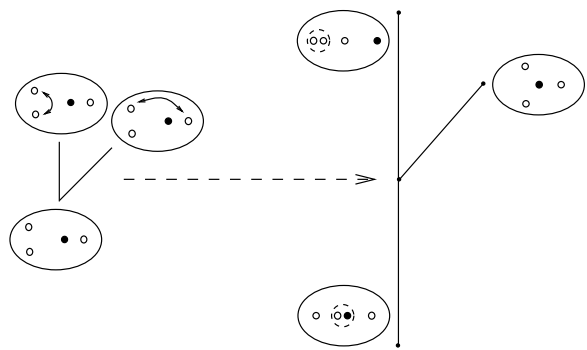}}
\end{picture}
\caption{\lb{4red-200}\protect\footnotesize\it The space of localized
4-vortex states is the ``Y'' shaped union of the two lines above. The
stable relative equilibrium~\rf{4red-30} is represented by the
endpoint of the line branching upwards towards the right. At the top and
bottom endpoints of the vertical branch are the $\pm\infty$ collision
states respectively. The bifurcation point in the middle is the
unstable relative equilibrium~\rf{4red-41}. From that point
proceeding upwards immediately yields to modes of motion corresponding to
the two connecting homoclinic orbits, as indicated on the left.}\end{figure}

The set~\rf{4red-32}, the top half of which is graphed on the left of
Figure~\rf{4red-35}, is homeomorphic to a two sphere and has three
singularities, corresponding to the two collision types of the reduced
phase space and to the stable relative equilibrium~\rf{4red-30}. If
the $\pm\infty$-collision states are removed then the phase space is a
cylinder; however, the symplectic form of this phase space is undefined
at one point, namely the stable equilibrium corresponding to the relative
equilibrium~\rf{4red-30}. An explicit map to this cylinder can be
constructed by deforming~\rf{4red-32} according to
\begin{equation}\lb{4red-37}
q_1\equiv p_1-\frac14(1-{p_2}^2),\quad q_2\equiv p_2,\quad q_3\equiv p_3,
\end{equation}
resulting in the surface plotted on the right of
Figure~\rf{4red-35}. The collision states
\begin{equation*}
(p_1,p_2,p_3)=(0,\pm1,0)
\end{equation*}
are unchanged but the result of the deformation uniquely intersects
radial half lines from the origin, and thus can be radially
homeomorphed to the sphere. Then, stereographically projecting from
the collision state $(q_1,q_2,q_3)=(0,-1,0)$ results in the plane
$\mbb C\setminus\{0\}$, and the pairing of the angle $\theta_{\mu_e}$
at the origin together with the logarithm $h_{\mu_e}$ of the radius
results in a map to the 2-torus such that the $\pm\infty$ energy
collision states are mapped to $\pm\infty$ respectively. The final
map to the cylindrical phase space $\{(h,\theta)\}$ is then
\begin{equation}\lb{4red-50}
h_{\mu_e}\equiv\frac12\ln\left(\frac{{q_1}^2+{q_3}^2}{q_2+
 \sqrt{{q_1}^2+{q_2}^2+{q_3}^2}}\right),\qquad
\theta_{\mu_e}\equiv\arctan\left(\frac{q_3}{q_1}\right),
\end{equation}
and~Figure~\rf{4red-40} shows the resulting phase portrait.

As stated at the beginning, the prime motivation of this article has
been the classification of 4-vortex states, produced through
collisions of two stable relative equilibria~\rf{4red-30}, with
momenta near $\mu_e\equiv(\Gamma\alpha^2/2,0)$ but far from the
relative equilibria~\rf{4red-30}. Such localized states ought to be
roughly regarded as orbits of the phase space in Figure~\rf{4red-40},
since there will be a fast time scale roughly corresponding to motion
along those orbits and the localized states will present external
properties corresponding to averages over that fast time scale. Thus,
the space of localized 4-vortex states corresponds to the quotient
space of the torus by the orbit equivalence relation of the phase space
of~Figure~\rf{4red-40}. This quotient is shown in Figure~\rf{4red-200}.

\frenchspacing


\begin{thebibliography}{}

\bibitem[\protect\citeauthoryear{Adams and Ratiu}{Adams and
  Ratiu}{1988}]{AdamsMRatiuTS-1988.1}
Adams, M. and T.~S. Ratiu [1988].
\newblock The three point vortex problem: commutative and non-commutative
  integrability.
\newblock In K.~R. Meyer and D.~G. Saari (Eds.), {\em Hamiltonian dynamical
  systems}, Volume~81 of {\em Contemporary Math.}, pp.\  245--257. AMS.

\bibitem[\protect\citeauthoryear{Aref and Pomphrey}{Aref and
  Pomphrey}{1982}]{ArefHPomphreyN-1982}
Aref, H. and N.~Pomphrey [1982].
\newblock Integrable and chaotic motions of four vortices {I}. {T}he case of
  identical vortices.
\newblock {\em Proc. Roy. Soc. London Ser. A\/}~{\em 380}, 359--387.


\bibitem[\protect\citeauthoryear{Eckhardt and Aref}{Eckhardt and
  Aref}{1988}]{EckhardtBArefH-1988}
Eckhardt, B. and H.~Aref [1988].
\newblock Integrable and chaotic motions of four vortices {II}. {C}ollision
  dynamics of vortex pairs.
\newblock {\em Philos. Trans. Roy. Soc. London Ser. A\/}~{\em 326}, 655--696.


\bibitem[\protect\citeauthoryear{Kidambi and Newton}{Kidambi and
  Newton}{1998}]{KidambiRNewtonPK-1998.1}
Kidambi, R. and P.~K. Newton [1998].
\newblock Motion of three point vortices on a sphere.
\newblock {\em Physica D\/}~{\em 116}, 143--175.


\bibitem[\protect\citeauthoryear{Patrick}{Patrick}{1992}]{PatrickGW-1992.1}
Patrick, G.~W. [1992].
\newblock Relative equilibria in {Hamiltonian} systems: {The} dynamic
  interpretation of nonlinear stability on the reduced phase space.
\newblock {\em J. Geom. Phys.\/}~{\em 9}, 111--119.


\bibitem[\protect\citeauthoryear{Patrick}{Patrick}{1995}]{PatrickGW-1995.1}
Patrick, G.~W. [1995].
\newblock Relative equilibria of {Hamiltonian} systems with symmetry:
  linearization, smoothness, and drift.
\newblock {\em J. Nonlin. Sc.\/}~{\em 5}, 373--418.


\bibitem[\protect\citeauthoryear{Patrick}{Patrick}{1996}]{PatrickGW-1993.1}
Patrick, G.~W. [1996].
\newblock Dynamics near stable relative equilibria at non-generic momenta: a
  numerical investigation.
\newblock {\em Fields Inst. Comm.\/}~{\em 8}, 127--142.


\bibitem[\protect\citeauthoryear{Patrick}{Patrick}{1999}]{PatrickGW-1998.1}
Patrick, G.~W. [1999].
\newblock Dynamics near relative equilibria: Nongeneric momenta at a 1:1
  group--reduced resonance.
\newblock {\em Math. Z.\/}~{\em 232}, 747--788.


\bibitem[\protect\citeauthoryear{Patrick}{Patrick}{2000}]{PatrickGW-1999.1}
Patrick, G.~W. [2000].
\newblock Dynamics of 4-vortex relative equilibria.
\newblock To appear {\it J. Nonlin. Sc.}

\bibitem[\protect\citeauthoryear{Pekarsky and Marsden}{Pekarsky and
  Marsden}{1998}]{PekarskySMarsdenJE-1998.1}
Pekarsky, S. and J.~E. Marsden [1998].
\newblock Point vortices on the sphere: stability of relative equilibria.
\newblock {\em J. Math. Phys.\/}~{\em 39}, 5894--5907.


\end{thebibliography}
\end{document}